\documentclass[review,12pt]{elsarticle}

\usepackage{amsmath,amssymb,amsfonts}
\usepackage{physics}
\usepackage[caption=false]{subfig}
\usepackage{graphicx}
\usepackage{setspace}
\usepackage{balance}
\def\BibTeX{{\rm B\kern-.05em{\sc i\kern-.025em b}\kern-.08em
    T\kern-.1667em\lower.7ex\hbox{E}\kern-.125emX}}

\newcommand{\ua}{\uparrow}
\newcommand{\nc}{\newcommand}

\nc{\da}{\downarrow} \nc{\hc}{\hat{c}} \nc{\hS}{\hat{S}}
\nc{\h}{\hat} \nc{\hT}{\h{T}} \nc{\rd}{\textrm{d}} \nc{\hR}{\hat{R}}
\nc{\tS}{\tilde{S}}\nc{\8}{\infty}\nc{\lgs}{\bra\ua,\phi|}\nc{\rgs}{|\ua,\phi\ket}
\nc{\hU}{\hat{U}}\nc{\lfs}{\bra\phi|}\nc{\rfs}{|\phi\ket}\nc{\hZ}{\hat{Z}}\nc{\hd}{\hat{d}}\nc{\mD}{\mathcal{D}}
\nc{\bd}{\bar{d}} \nc{\bc}{\bar{c}} \nc{\mc}{\mathcal} \nc{\mG}{\mathcal{G}}

\journal{GLOBECOM 2022}

\begin{document}

\begin{frontmatter}

\title{Quasi-Chaotic Oscillators Based on\\Modular Quantum Circuits
}

\author{Andrea Ceschini}
\ead{andrea.ceschini@uniroma1.it}
\author{Antonello Rosato}
\ead{antonello.rosato@uniroma1.it}
\author{Massimo Panella\corref{cor1}}
\ead{massimo.panella@uniroma1.it}

\cortext[cor1]{Corresponding author}

\address{Department of Information Engineering, Electronics and Telecommunications\\
				 University of Rome ``La Sapienza'', Via Eudossiana 18, 00184 Rome, Italy.}

\begin{abstract}
Digital circuits based on residue number systems have been considered to produce a pseudo-random behavior. The present work is an initial step towards the complete implementation of those systems for similar applications using quantum technology. We propose the implementation of a quasi-chaotic oscillator based on quantum modular addition and multiplication and we prove that quantum computing allows the parallel processing of data, paving the way for a fast and robust multi-channel encryption/decryption scheme. The resulting structure is assessed by several experiments in order to ascertain the desired noise-like behavior.
\end{abstract}


\end{frontmatter}

\section{Introduction}
\label{sec:intro}
Quasi-Chaotic (QC) generators represent a particular class of digital systems with a range of implementations in different sectors \cite{RNS_Book}. In particular, QC generators are extremely suitable for encryption and for encoding/decoding signals for secure communication \cite{RNS_App}. Their architecture can also be tailored to secure communication systems able to self-correcting transmission errors \cite{RRNS_Panella}. Therefore, QC generators are considered particularly suitable to exploit the potentiality of discrete-time circuits in the area of secure and covert data transmission. In the past, residue number system (RNS) architectures have been proposed to implement QC generators \cite{RNS_Panella1}, as they make use of modular arithmetic by which the pseudo-random behavior can be obtained in a straightforward manner and with interesting properties regarding VLSI deploying, modularity, speed, fault tolerance and low-power consumption \cite{RNS_Panella2}. The flexibility of the RNS approach allows to easily met all the major requirements related to secure communication systems. In fact, RNS is the realization of an unweighted number system where numbers are represented using a set of relatively prime positive integers called moduli, which are used to perform specific arithmetic operations independently of other moduli. This function enables parallel processing on all channels without transmission to other channels.

\begin{figure}[!ht]
\centering{\includegraphics[width=0.45\columnwidth]{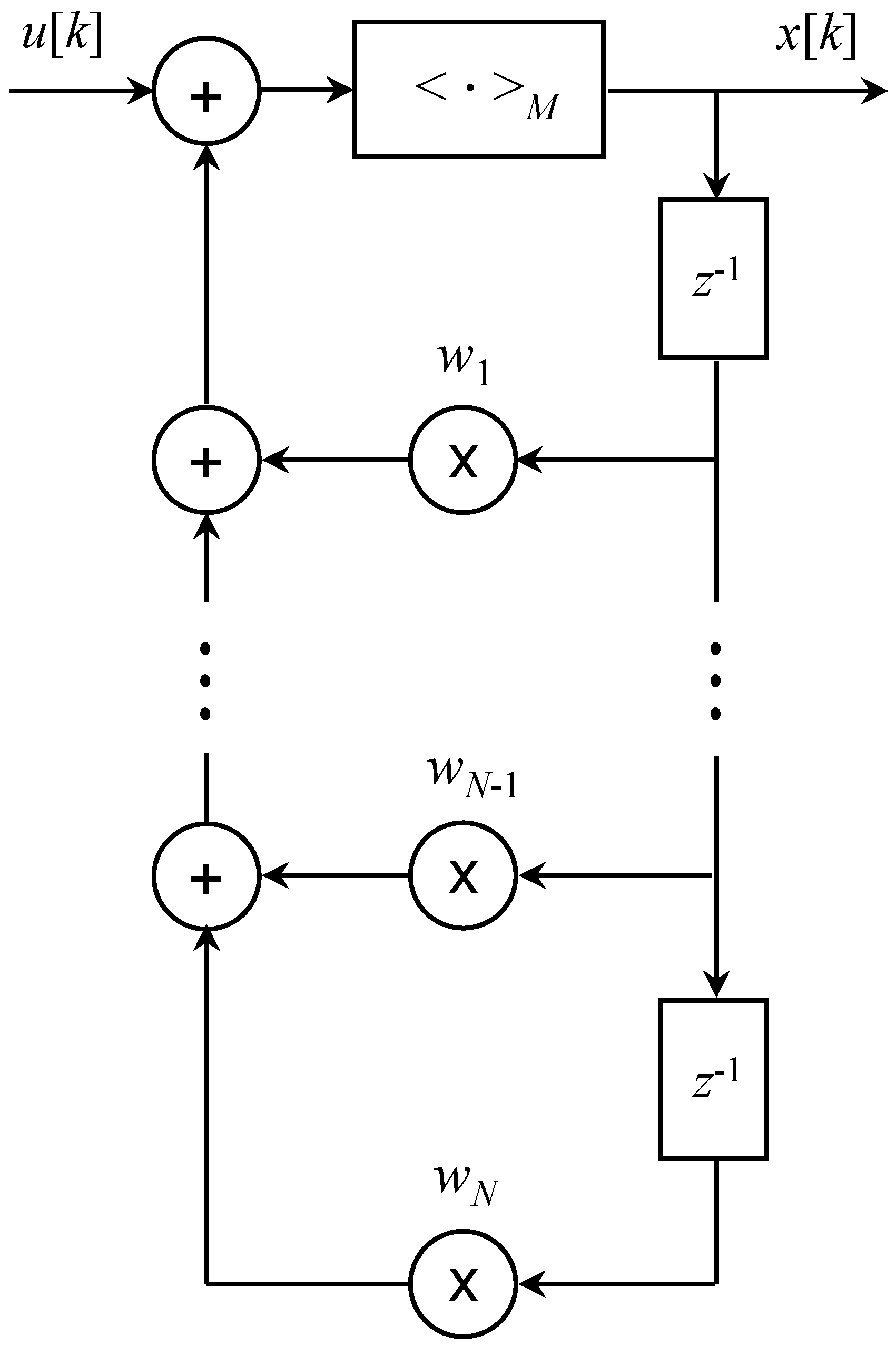}}
\caption{Schematic diagram of a modular QC generator. Despite only one modulo operation is reported, the latter can be distributed over all sums and multiplications in the filter.}
\label{fig:IIR}
\end{figure}

In this framework, a QC generator can be implemented by means of the nonlinear IIR filter shown in Fig.~\ref{fig:IIR}, which is characterized by the following difference equation:
\begin{equation}
x[k]=\left\langle {u[k]+\sum_{i=1}^N{w_i\,x[k-i]}} \right\rangle_M ,
\label{eq:IIR}
\end{equation}
where all algebraic operations are defined modulo $M$ and all elements of input and output time series, as well as the filter coefficients $w_i$, $i=1\dots N$, belong to the ring $R(M)$ of the integers modulo $M$. Usually, $u[k]$ is the input time series to be encrypted, while $x[k]$ is the resulting encrypted version; the $N$-order IIR QC generator is characterized by the set of coefficients $w_i$, which can be considered as the encryption key for secure communication \cite{dachselt_1998}.

The inverse FIR system that allows the decoding of the modulated input $u[k]$ from the received sequence $\tilde{x}[k]$ is obtained by the following difference equation:
\begin{equation}
\tilde{u}[k]=\left\langle {\tilde{x}[k]-\sum_{i=1}^N{w_i\,\tilde{x}[k-i]}} \right\rangle_M ,
\label{eq:IIR_2}
\end{equation}
where $\tilde{u}[k]$ is the decoder output, i.e. the decoded time series relative to $u[k]$, while the decoder input $\tilde{x}[k]$ is the sequence received on the other side of a communication channel. For the sake of simplicity, given the symmetric nature of any encoder/decoder pair, only the encoder system is considered in the present paper. In other words, the involved algebraic operations are the same and hence, the quantum implementation of both systems follows the same main rules.

In order to measure the strength of the QC modulation obtained by \eqref{eq:IIR}, the noise-like behavior of the zero-input response $x[k]$, $k \geq 0$, is measured when $u[k]$ is set to zero and the IIR filter is used as a QC oscillator with given initial conditions on $x[-1]$, $x[-2]$, up to $x[-N]$:
\begin{equation}
x[k]=\left\langle {\sum_{i=1}^N{w_i\,x[k-i]}} \right\rangle_M .
\label{eq:zero_input}
\end{equation}

The advent of quantum communication can become an enabler for the development of new cryptography techniques. In particular, QC generators have the potential to be at the core of this development. To the best of our knowledge, nobody has so far implemented or discussed a quantum implementation of a nonlinear IIR filter acting as a QC generator. In light of this, the solution proposed in the present paper aims at designing a quantum QC oscillator through modular quantum arithmetic operations. Despite the technological issues that still affect real quantum systems \cite{preskill_2021}, we provide a preliminary circuit implementation of a QC system in the quantum framework.

QC generators could benefit from quantum computing in several ways, by using parallel encryption/decryption schemes \cite{nielsen_2010} or even adopting computationally efficient neural networks via quantum superposition \cite{salamat_2018,chervyakov_2020}. As a matter of fact, while classical computers are based on bits as elementary units of information, with mutually exclusive values of 0 and 1, quantum devices use qubits as building blocks of the computation: the 0 and 1 states coexist simultaneously in a probabilistic superposition. The latter property allows quantum computers to process data in parallel in a high-dimensional form with very few qubits \cite{weigold_2021}, with the goal of adopting the quantum QC generators proposed herein for encryption/decryption in high-speed secure communication channels.

The challenge faced in this paper is to replicate the modular operations in \eqref{eq:IIR} with suited quantum gates, in order to obtain a QC behavior given a given initial condition. The correct functioning of the proposed quantum QC circuit is experimentally validated with the help of Qiskit\textsuperscript\texttrademark, which is a framework for quantum circuits' simulation developed in Python. Taking a single section of the filter, all of the possible input-output pairs obtained with the quantum circuit are compared to the desired result. Successively, the zero-input response of the whole quantum time series is validated against the classical counterpart and the effects of bit-flip measurement errors are studied for different noise levels. Finally, the autocorrelation function of the zero-input response is calculated to verify its similarity to an uncorrelated noise sequence.

\section{Quantum Implementation of QC Oscillators}
\label{sec:implementation}
\begin{figure}[!ht]
 \centering{\includegraphics[width=0.5\columnwidth]{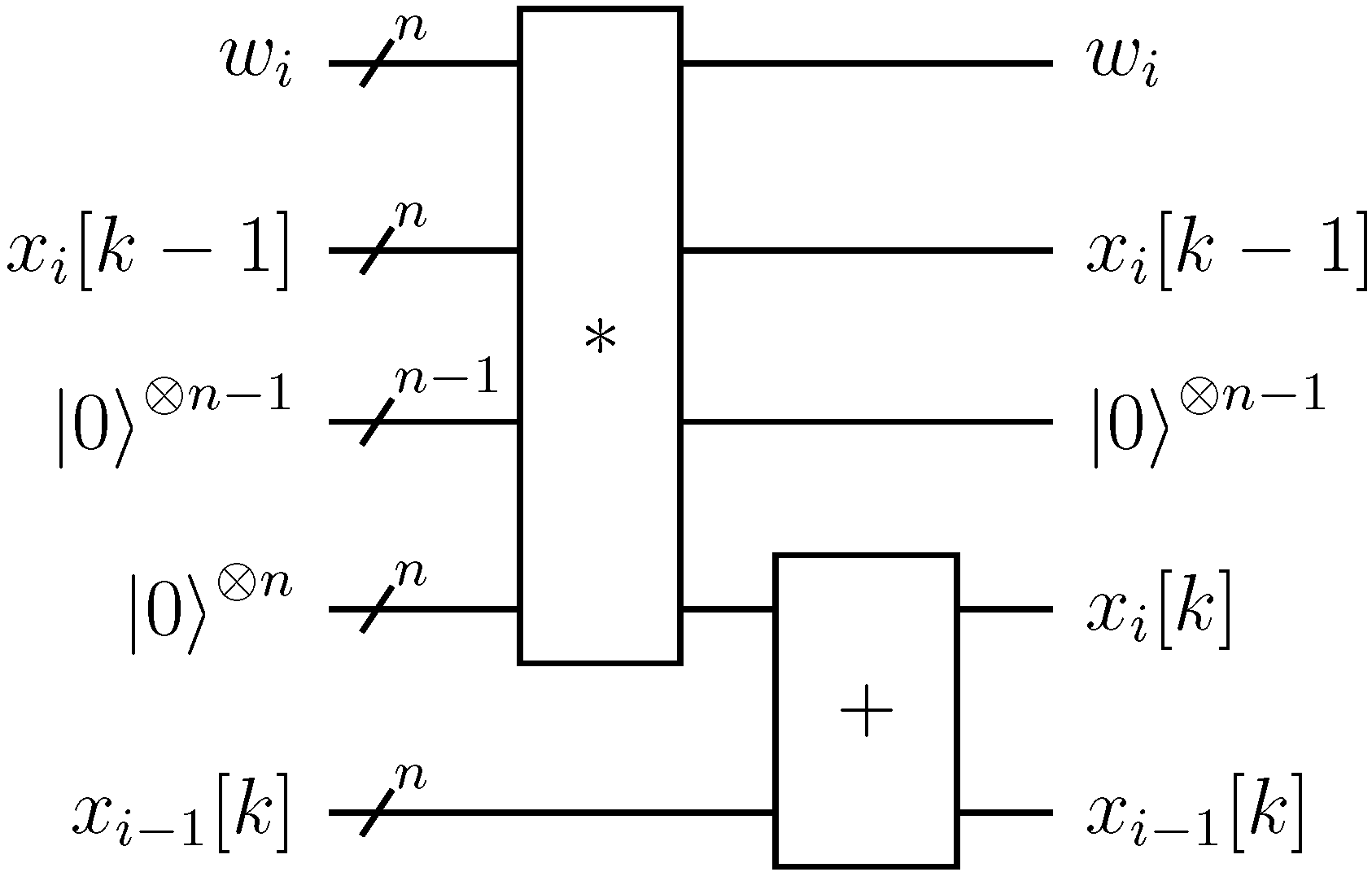}}
		\vspace{3pt}
    \caption{A snapshot of the $i$th section of the quantum QC filter associated with the $i$th delay of the IIR structure.}
    \label{fig:QQC}
\end{figure}

From a structural perspective, there are no substantial modifications to be made to the architecture based on CMOS VLSI technology, as the one of a quantum QC oscillator should be equivalent to the classical counterpart. It can be designed via a series of modular quantum arithmetic functions such as quantum addition and quantum multiplication. Recalling the architecture in Fig.~\ref{fig:IIR}, the design of a single section of a quantum QC filter is illustrated in Fig.~\ref{fig:QQC}.

Given a sequence of $P$ samples, each $i$th section of the quantum filter, ${i = 1\dots N}$, is composed as follows: $x_{i-1}[k]$ is the input to the $i$th section, $w_{i}$ is the $i$th corresponding coefficient, $x_i[k-1]$ is a first-order delay and $x_i[k]$ is the output of the filter. Since quantum computation must be reversible, for each iteration $i$ relative to sample $k$ of the filter, the inputs $w_{i}$, $x_{i-1}[k]$ and $x_{i}[k-1]$ are also present in the output. Moreover, some ancillas initialized to $\ket{0}$ act as temporary registers to store intermediate steps of the algorithm. A part of the ancillas is used to encode the output $x_{i}[k]$, while the rest is restored to zero via uncomputing \cite{johnston_2019}, in order to use it for further calculations. In fact, modern quantum systems are able to provide just a small number of logical qubits, therefore it is of paramount importance to save and reuse as many qubits as possible during the computation.

Considering that we are dealing with binary representations of integer numbers, working with $n$-qubit binary strings directly leads to a modular arithmetic representation with an $M=2^n$ modulo; thus, the range of possible values is $[0, 2^n-1]$. In addition, given two numbers $a$ and $b$, the following property of modulo $n$ arithmetic is exploited:
\begin{equation}
    {( a\, \operatorname{op}\, b )\,\operatorname{mod}\, n} = ((a\, \operatorname{mod}\, n) \,\operatorname{op}\, (b \,\operatorname{mod}\, n) )\, \operatorname{mod}\, n \,,
\end{equation}
for any operation $\operatorname{op}\in\{+, -, * \}$ and where ${(\cdot)\,\operatorname{mod}\,n}$ denotes modulo $n$ residue extraction.
Binary arithmetic intrinsically leverages this property to handle the overflow with a $2^n$ modulus reduction. 

The elementary operations computed inside a quantum QC generator are analogous to the ones in the classical circuit. A quantum version of modular addition, bit shift and modular multiplication is hereafter presented.

\subsection{Quantum Modular Addition}
\begin{figure}[!ht]
   \centering{\includegraphics[width=0.93\columnwidth]{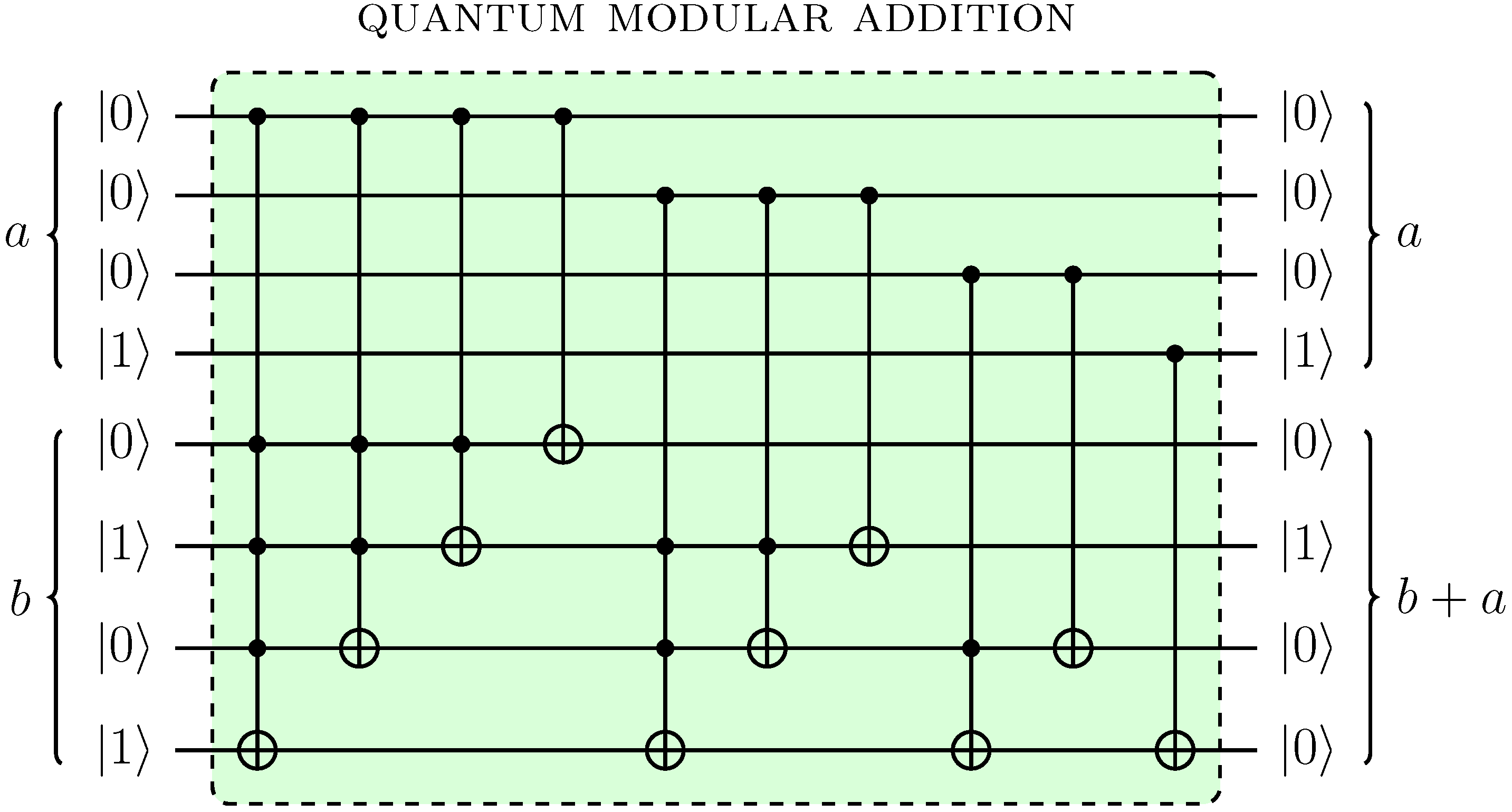}}
		\caption{Quantum modular addition between two numbers \textnormal{a} and \textnormal{b} in modulo $2^4 = 16$. The least significant qubit is on top.}
    \label{fig:QADD}
\end{figure}

Quantum addition between two numbers $a$ and $b$ can be thought as a bit-wise increment of $b$ controlled by the corresponding qubits of $a$, as depicted in Fig.~\ref{fig:QADD}. This operation does not need ancillary registers to store the result, because the value of the first addend is added to the second one in an addition assignment fashion. The sum is computed modulo $n$ due to the possible overflow. In order to deal with quantum principles, the $+=$ operator is implemented instead; it directly adds one number onto another.

\subsection{Quantum Bit Shift}
\begin{figure}[!ht]
		\centering{\includegraphics[width=0.65\columnwidth]{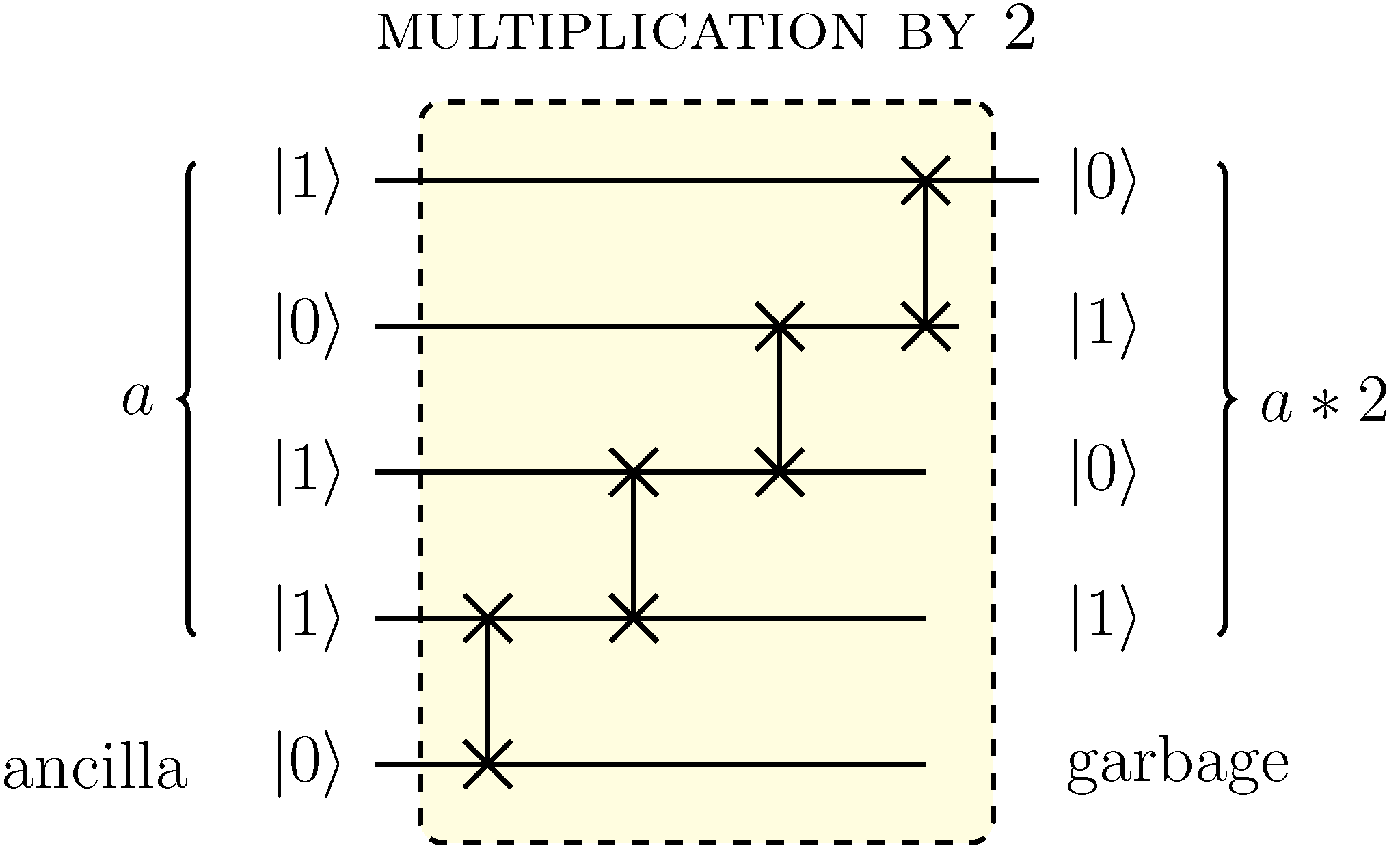}}
    \caption{Quantum multiplication by 2 using left shifting.}
    \label{fig:SWAP}
\end{figure}

Bit shift is used to multiply or divide a binary number by powers of 2. In particular, a single left shift corresponds to a multiplication by 2: all the bits are shifted off to the left, so that the most significant bit is discarded and a $0$ bit is inserted at the end of the binary string. In a quantum framework, left shift can be carried out with a series of `swap' gates in cascade and an ancilla qubit initialized to $\ket{0}$, as shown in Fig.~\ref{fig:SWAP}. The swap gates are responsible for moving the qubits left. The ancilla ends up storing the most significant qubit, while the $\ket{0}$ previously contained in the ancilla is inserted at the end of the sequence. The unused output qubit at the end of the computation is a `garbage'.

\subsection{Quantum Modular Multiplication}
\begin{figure*}[!ht]
 \hspace{-3cm}{\includegraphics[width=1.45\textwidth]{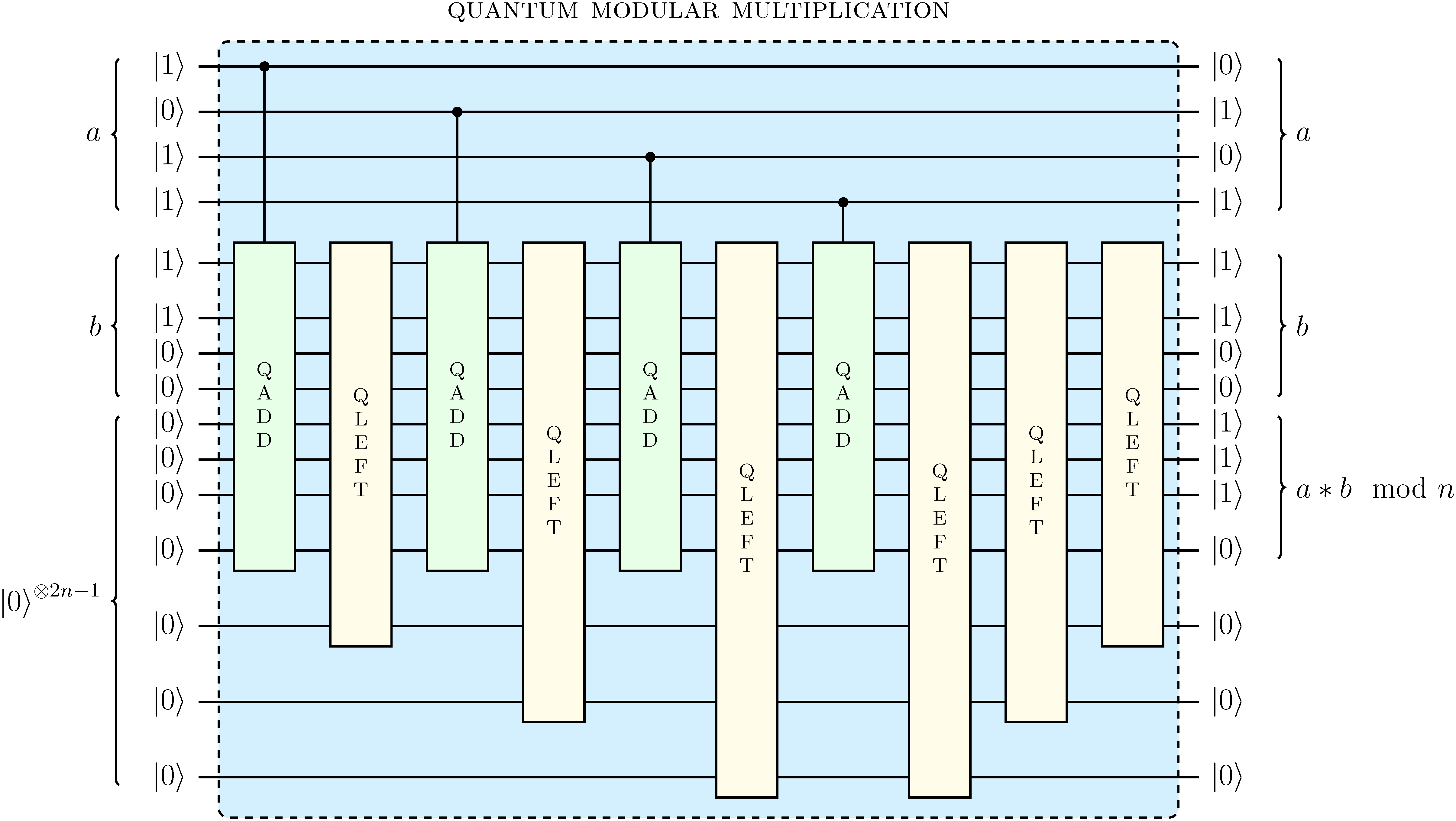}}
    \caption{An example of 4-qubit quantum multiplier based on a sequence of controlled additions and left shifts. The QADD and QLEFT transformations correspond to quantum modular addition and left shift, respectively.}
    \label{fig:MULT}
\end{figure*}

A quantum modular multiplier can be constructed using a combination of controlled modular adders and left shifters, as proposed in \cite{rines_2018}. The sequence of controlled additions accumulates into a product register, so that the final product ends up being already reduced modulo $n$. Given two binary numbers $a$ and $b$ of length $n$, each partial product is of the form ${a_i(2^ib) \mod n}$, ${i = 0\dots (n-1)}$. Each addition in the multiplier uses a value that is twice the previous value, therefore we just need to shift the value by one position for each addition. 
An example of quantum modular multiplication circuit is represented in Fig.~\ref{fig:MULT}. In order to free some ancilla qubits at the end of the computation and to retrieve all the factors of the multiplication to their original state, uncomputing is employed. It reverses the operations that entangled such qubits by applying the same quantum gates in a reversed order. This way, by reversing the left shifts transformations, the registers at the output which do not contain the multiplication result are restored to their initial value.

\subsection{Unrolled Quantum QC Oscillator}
\begin{figure}[!ht]
		\centering{\includegraphics[width=0.9\columnwidth]{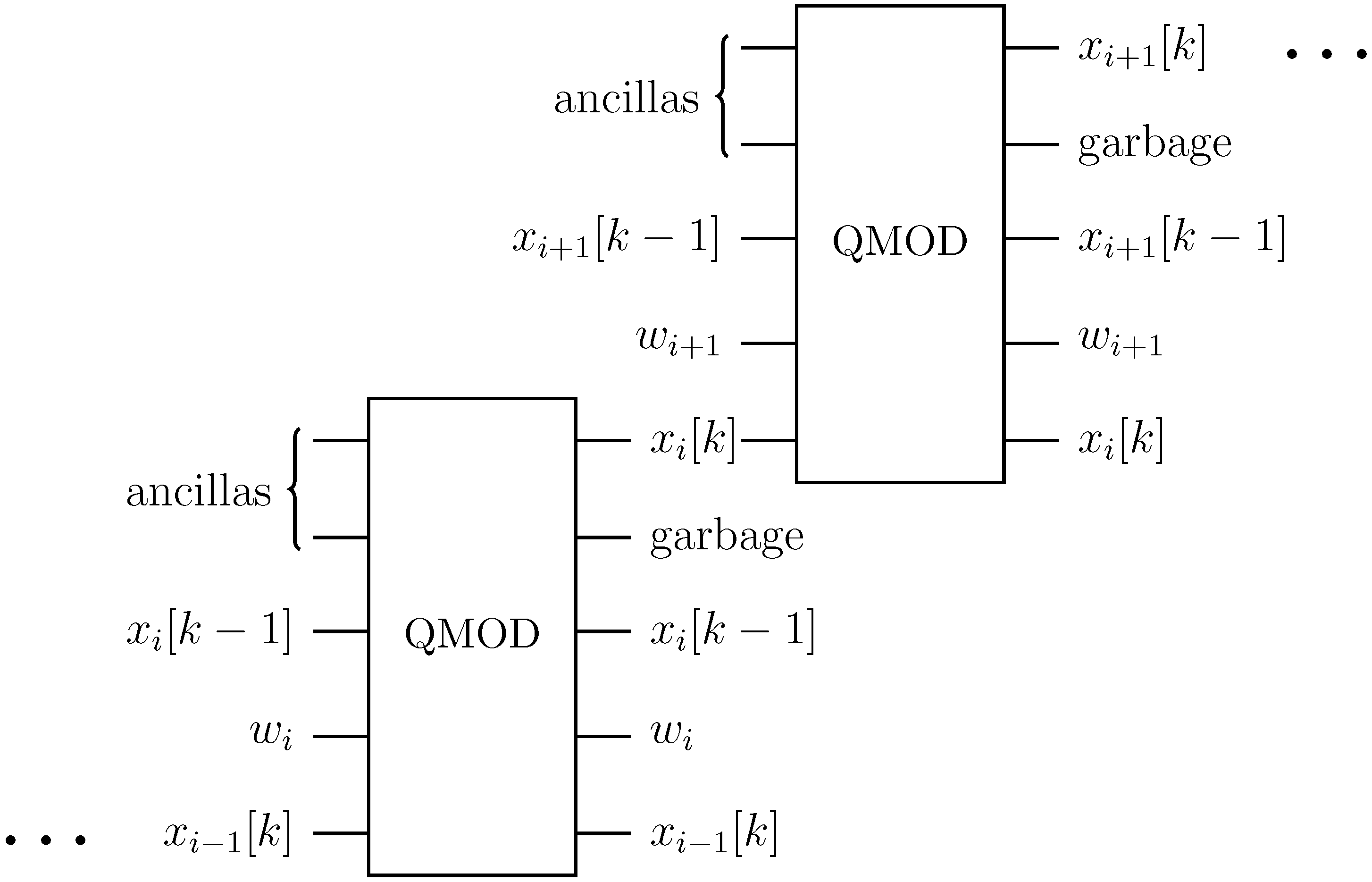}}
    \caption{Unrolled scheme of a quantum QC oscillator. Each single section is denoted as QMOD.}
    \label{fig:UNROLLED}
\end{figure}

The architecture of a quantum QC oscillator is depicted in Fig.~\ref{fig:UNROLLED}. It is an unrolled concatenation of quantum circuits from Fig.~\ref{fig:QQC}, with a cascade of $N$ first-order quantum RNS sections. The output $x_i[k]$ of the $i$th section becomes the input of the next $i + 1$th section. The signals are not stored in external registers due to the no-cloning theorem, but they are processed in a forward manner. Input qubits could be put into superposition to initialize the filter with every possible combination of inputs; successive processing of the resulting output may be performed to gain quantum advantage.

\subsection{Complexity Analysis of the Quantum QC Circuit}
An in-depth analysis of the proposed quantum QC generator is hereafter illustrated. First, a qubit count is discussed to analyze the spatial complexity of the circuit. Every $i$th QMOD section of the filter receives at the input $n$ qubit strings for the coefficient $w_i$, $n$ qubit strings for the input $x_{i-1}\text{[}k\text{]}$, $n$ qubit strings for the first-order delay $x_{i}\text{[}k-1\text{]}$ and $2n-1$ ancillary qubits. In total, $5n-1$ qubits are needed to perform a QMOD operation. However, at the output, ${n-1}$ ancillary qubits can be reused in the computation since they are restored via uncomputing. Moreover, the filter response $x_{i}\text{[}k\text{]}$ is also reused in the next ${i+1}$ section of the filter. Based on these considerations, the subsequent QMOD sections in a pipeline architecture will only need $3n$ qubits each. Therefore, the final qubit count $C(k)$ for $k$ samples in a quantum QC generator is given by the following function:
\begin{equation}
C(k)= \begin{cases}
    5n-1,& k = 1\\
    5n-1 + 3nk,              & k\geq 2
\end{cases}
\label{eq:spatial_complexity}
\end{equation}
and hence, the complexity is linear with respect to $n$.

As for the spatial complexity, the quantum cost in terms of primitive gates is strictly related to the size $n$ of the input strings. Every QMOD section has the same number of gates and is composed of a quantum modular addition QADD and a quantum modular multiplication QMUL. For the sake of simplicity, let us consider all the qubits connected to each other and a depth of $1$ for a generic $n$-Toffoli gate. The QADD operation has a depth proportional to $\mathcal{O}(n^2)$. The QMUL operation is a sequence of $n$ QADD and $2n-1$ QLEFT operations, therefore its depth is in the order of $\mathcal{O}(n^3)$. As a result, the overall depth of a single QMOD section of the filter is in the order of $\mathcal{O}(n^3)$.

\section{Experimental Analysis}
\label{sec:behavior}
We performed three different experiments to validate our proposed quantum approach. All the tests were conducted using Qiskit's 32-qubit simulator; each input string was represented in the range $[0, 15]$ using $4$ qubits in modulo $M=16$ arithmetic. The choice of using only $4$ qubits to encode binary numbers was due to actual technological constraints. This was also the rationale behind the choice of adopting a $2^n$ modulo for each section of the filter, which is easy to compute through simple overflow. Further experiments with an arbitrary modulo may be performed in future works, where the number of available qubits will no more be a constraint and fault-tolerance quantum computers will be more widely available.

\subsection{Experiment A}
\begin{figure}[!ht]
\centering{\includegraphics[width=0.95\columnwidth]{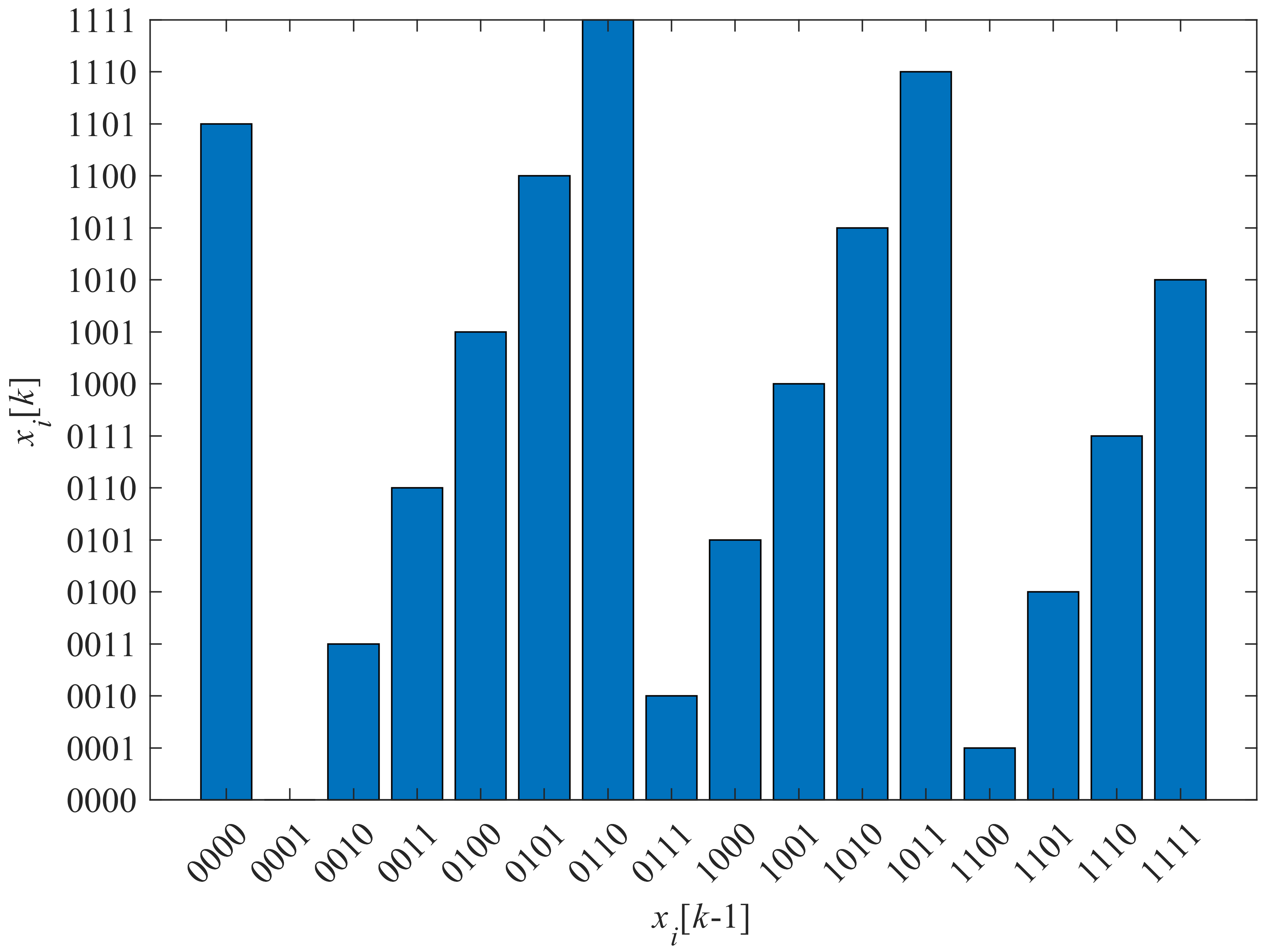}}
\caption{The output $x_i[k]$ of a single section of the quantum circuit with respect to different delays $x_i[k-1]$.}
\label{fig:EXP1}
\end{figure}

A single section of the quantum QC oscillator was evaluated. The behavior of such quantum circuit exactly reflected the response obtained by its classical counterpart for every possible combination of input string. Without loss of generality, only a test with a fixed coefficient $w_i = \ket{0011} = 3$ and a constant input $x_{i-1}[k] = \ket{1101} = 13$ is hereafter reported. The only parameter that varied was the delay $x_i[k-1]$, which ranged between $0$ and $15$. The correct functioning of such single QMOD section is presented in Fig.~\ref{fig:EXP1}.

\subsection{Experiment B}
\begin{figure}[!ht]
    \subfloat[\label{fig:comparison_quantum}]{%
      \includegraphics[width=0.95\columnwidth]{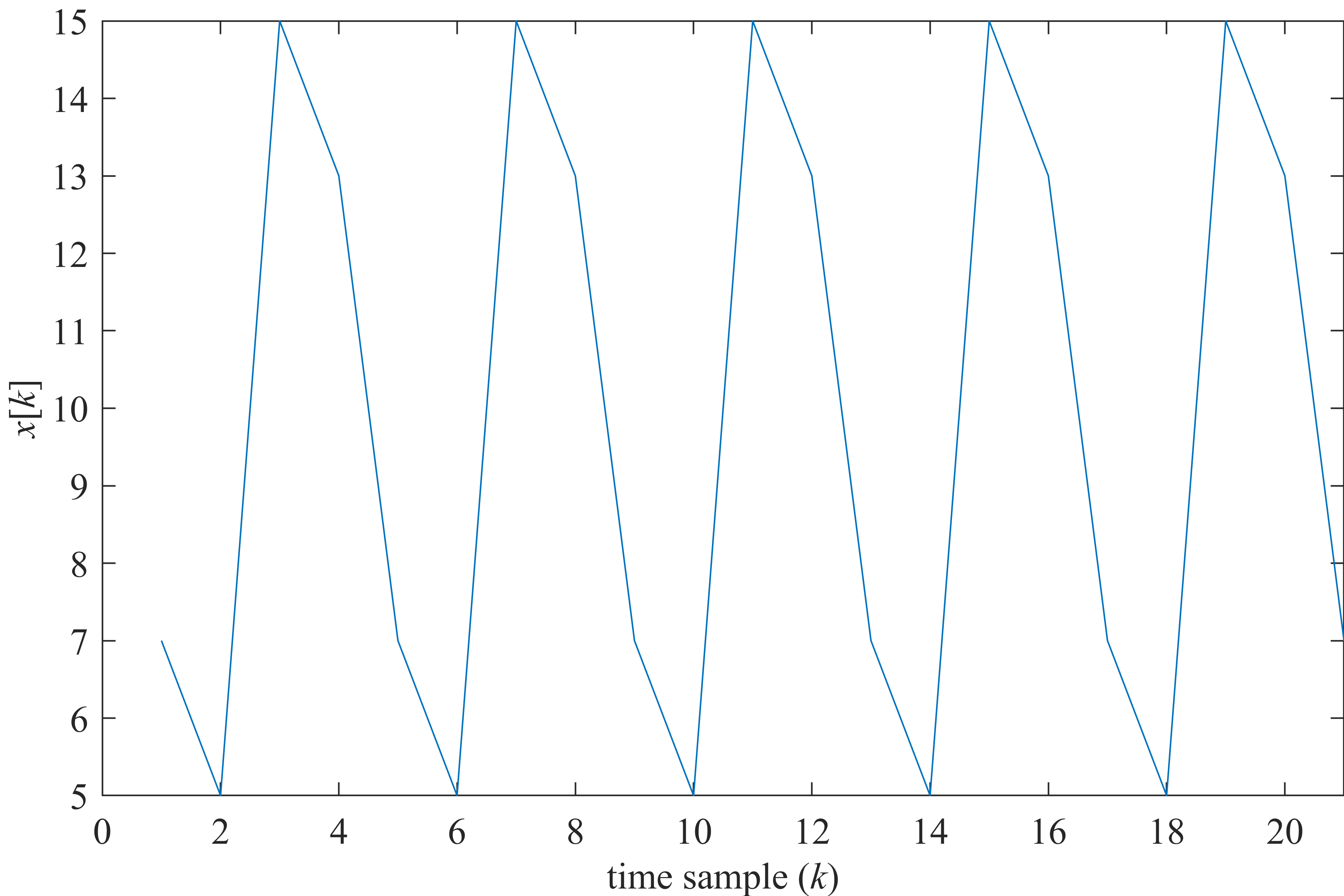}
    }
    \\[6pt]
    \subfloat[\label{fig:comparison_VLSI}]{%
      \includegraphics[width=0.95\columnwidth]{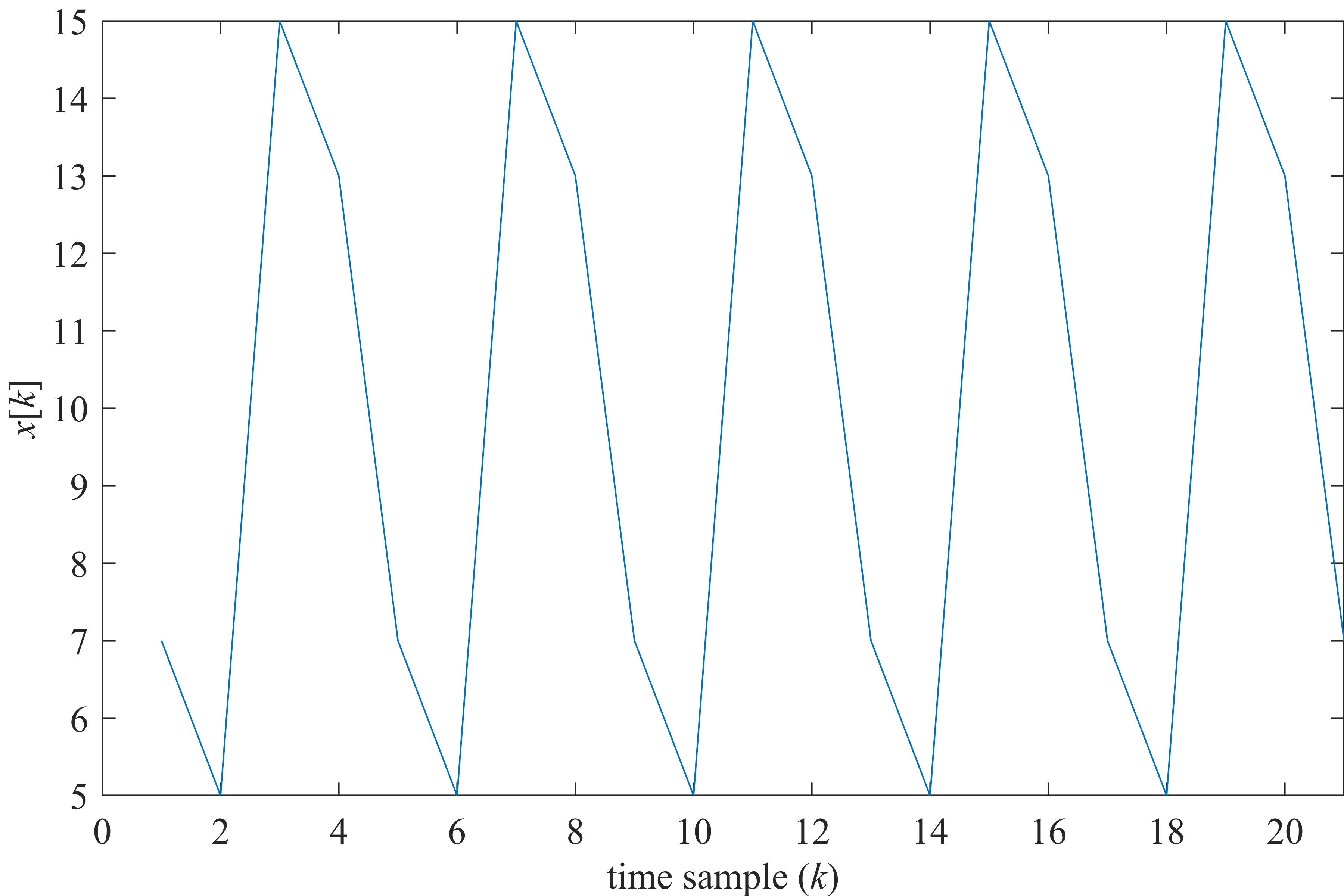}
    }
    \caption{Comparison between zero-input responses: (a) ideal quantum QC oscillator; (b) classical QC oscillator based on VLSI technology. Only 21 out of 50 samples are reported for the sake of illustration.}
    \label{fig:EXP2}
  \end{figure}

A quantum QC generator with filter order $N = 1$ was realized to calculate the zero-input response for ${k=1 \dots 50}$ samples. Given the initial conditions of $w_1 = \ket{0011} = 3$, $x_1[0] = \ket{1101} = 13$, $x_0[1] = \ket{0000} = 0$, $M = 16$, the output $x_1[k]$ of the circuit was calculated and validated against the classical counterpart, calculated in Python instead. To overcome the constraints on the number of qubits in the simulator, a single QMOD section was implemented at each step rather than the entire architecture, which was too demanding in terms of resources. For every new sample $k$, the input parameters were adjusted accordingly with appropriate quantum gates. At the end of the QMOD computation, the qubits were reset to reuse them in the next step of the algorithm. As demonstrated by Fig.~\ref{fig:EXP2}, the behavior of the quantum QC generator in an ideal setting, i.e. without considering noise effects, perfectly overlaps to the curve originated by the classical circuit. 

\begin{figure}[!ht]
\centering{\includegraphics[width=0.95\columnwidth]{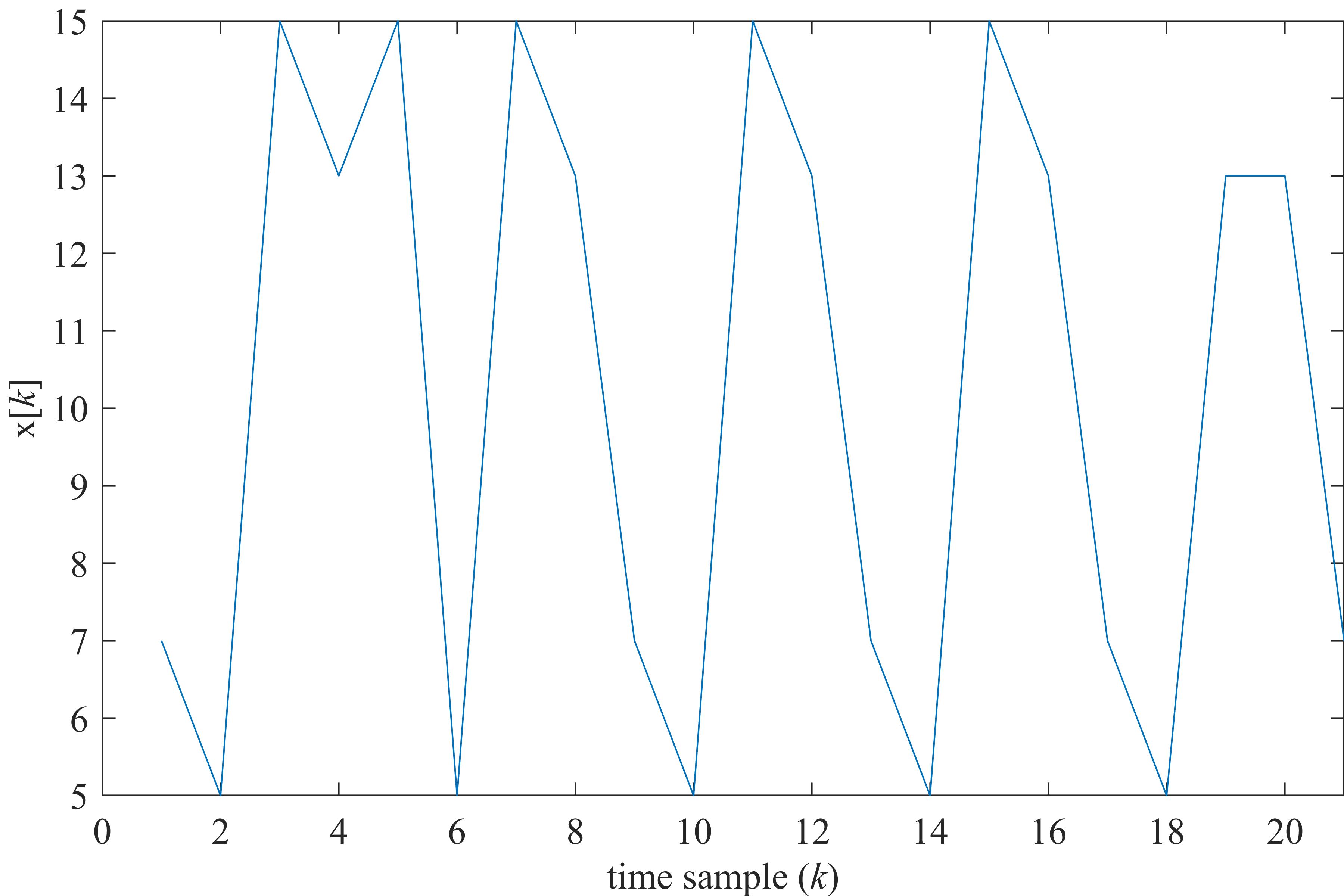}}
\caption{Zero-input response with a bit-flip measurement error having probability $p=0.02$.}
\label{fig:EXP2_2}
\vspace{-6pt}
\end{figure}

\begin{figure}[!ht]
\centering{\includegraphics[width=0.95\columnwidth]{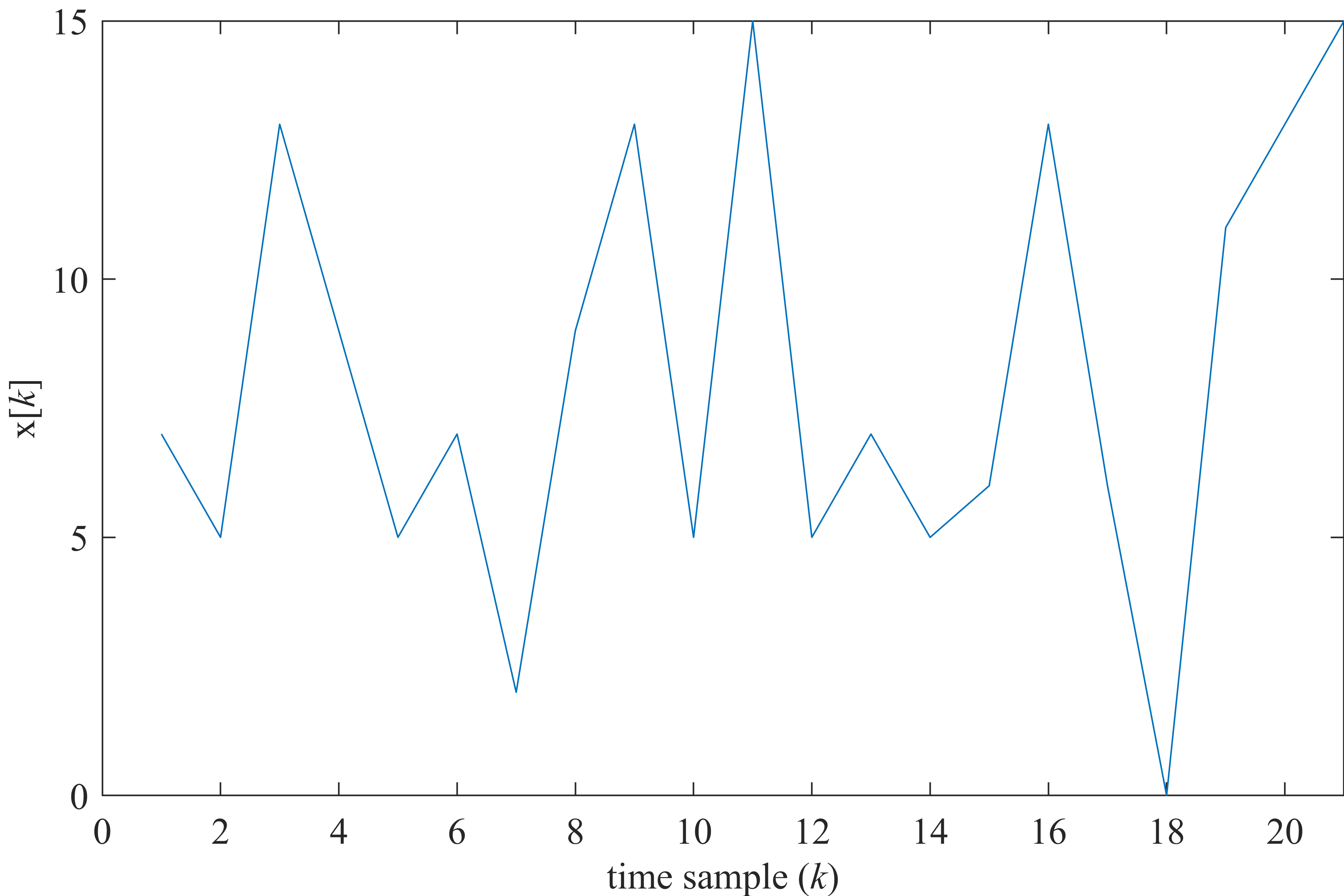}}
\caption{Zero-input response with a bit-flip measurement error having probability $p=0.16$.}
\label{fig:EXP2_3}
\end{figure}

\begin{figure}[!ht]
\centering{\includegraphics[width=0.95\columnwidth]{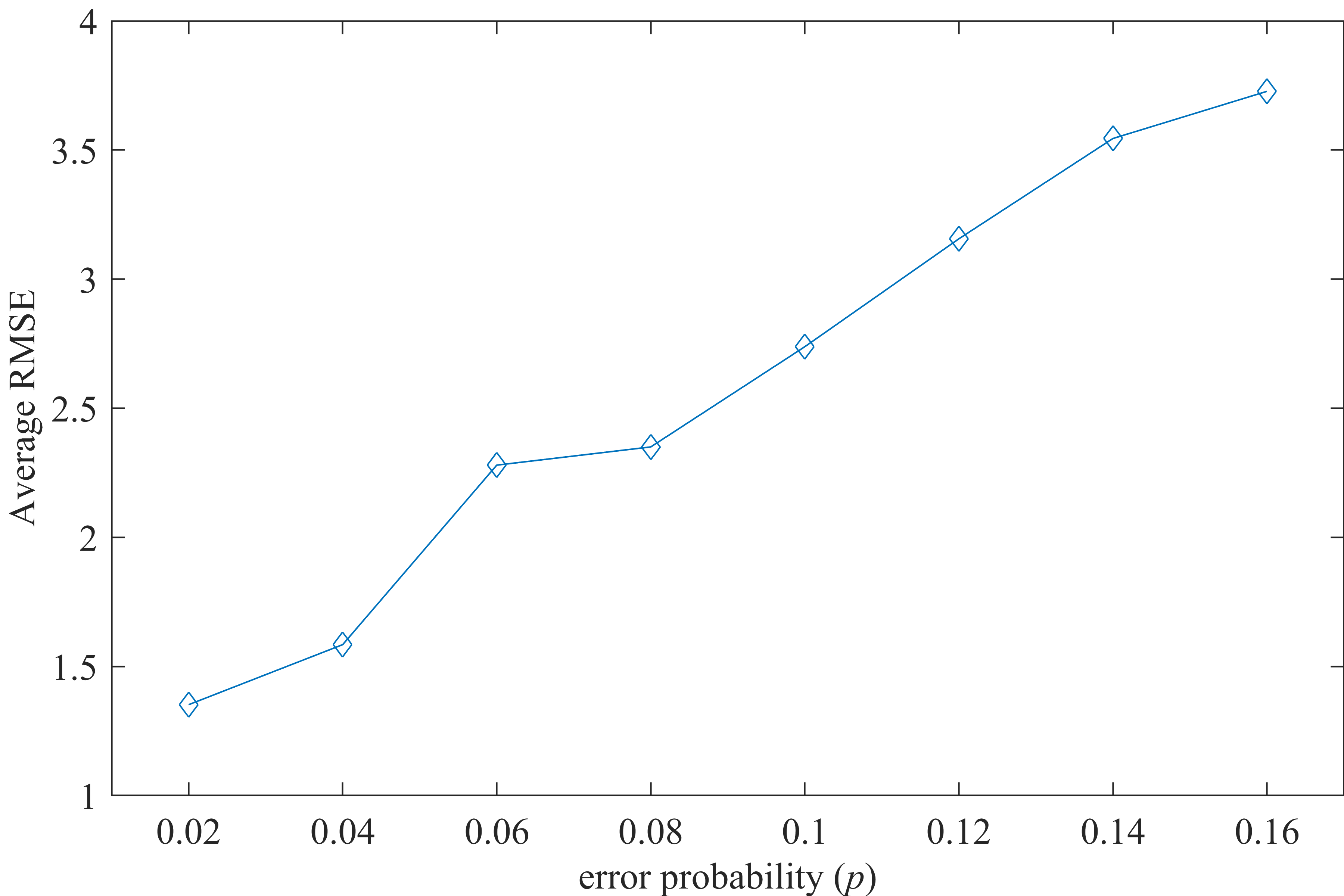}}
\caption{Average RMSE of zero-input response versus different noise levels.}
\label{fig:EXP2_4}
\vspace{3pt}
\end{figure}

In order to evaluate the robustness of the proposed quantum circuit in presence of several noise levels, we also performed the experiment after injecting bit-flip measurement errors to all the qubits, which is a common case study in quantum information theory research. Bit-flip measurement errors consist in flipping the state of a qubit with probability $p$ during a measurement. For $p=0.02$, the behavior of the quantum circuit still follows the expected output, as shown in Fig.~\ref{fig:EXP2_2}. 
For increasing values of probability $p$, the noise level progressively impairs the circuit's output; in the case $p=0.16$, the final outcome becomes completely dissimilar from the desired result as illustrated in Fig.~\ref{fig:EXP2_3}.
An overview of the Root Mean Squared Error (RMSE) with respect the ideal behavior, averaged over 10 different runs, is represented in Fig.~\ref{fig:EXP2_4} against the noise level $p$. 
As the probability of occurring in bit-flip measurement errors increases, the RMSE increases accordingly and almost linearly.

\subsection{Experiment C}
\begin{figure}[!ht]
\centering{\includegraphics[width=0.95\columnwidth]{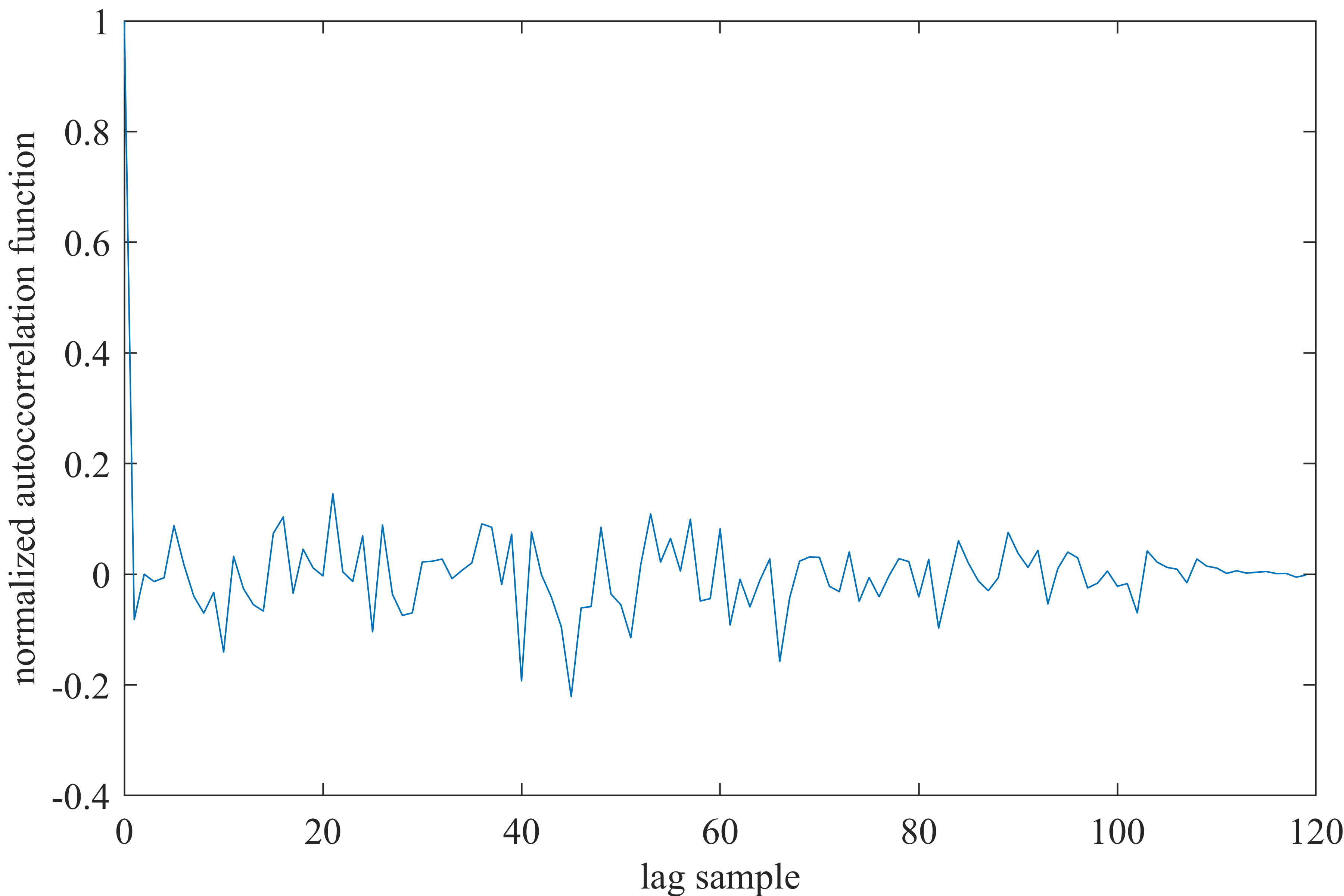}}
\caption{Normalized autocorrelation of the zero-input response with period 120 samples, obtained with $M=16$ and filter coefficients ${w_1=13}$, ${w_2=w_3=0}$, ${w_4=1}$.}
\label{fig:autocorr}
\end{figure}

Strong properties of QC generators are demonstrated when data belong to a Galois field GF($M$) defined by a prime modulus $M$ \cite{RNS_Panella1}. In case of even moduli of type ${M=2^n}$, weakened properties pertaining to a shorter periodicity of time series and to their dependence on initial conditions are balanced by the increased efficiency on hardware implementations. However, the autocorrelation of the zero-input response for the considered quantum QC oscillators results similar to that of an uncorrelated noise sequence, as proved in Fig.~\ref{fig:autocorr} in the case $M=2^4$ ad $N=4$. 

The choice of the modulus $M$ is constrained by the characteristics of the input-output signals, when the RNS system is applied to a digital signal processing application. For instance, it is related to the resolution (i.e., the number of levels) by which digital signals are represented. Usually, input-output signals are considered in a signed form onto a dynamic range of the modular representation: ${[-(M-1)/{2},\,(M-1)/{2}]}$ if $M$ is odd; ${[-{M}/{2},\,{M}/{2}-1]}$ if $M$ is even. Each integer in the dynamic range is mapped onto the legitimate range ${[0,\,M-1]}$, which represents the actual computational range of the RNS system. For instance, ${[-(M-1)/{2},\,-1]}$ maps onto ${[(M+1)/{2},\,M-1]}$ if $M$ is odd, whereas ${[-{M}/{2},\,-1]}$ maps onto ${[{M}/{2},\,M-1]}$ if $M$ is even.

\section{Conclusion}
In this work, we presented a study on quasi-chaotic generation based on quantum modular arithmetic. We were able to experimentally validate the proposed approach with different simulations, assessing the performance with respect to the original VLSI solution. As for practical applications, we argue that quantum superposition would allow to parallelize the execution of a QC generator for different encryption keys, providing an exponential number of filter responses simultaneously. Accordingly, quantum QC oscillators should be employed as part of a larger algorithm, selecting just one result among the possible outputs based on arbitrary criteria or an oracle function could be applied to the outputs in superposition while adopting some amplitude amplification procedures.

Rather than a specific solution for all the inputs in superposition, some derived properties of the overlapped output may be considered, such as its sum, the minimum value or the presence of a certain element. Future works might also investigate on extended implementations dealing with modulo $M$ arithmetic, where $M$ is a prime number ensuring a stronger noise-like behavior. In this regard, further experiments would benefit from running on actual quantum hardware to underpin the practical advantages of a quantum QC oscillator.

\end{document}